\documentclass[prl,aps,twocolumn,superscriptaddress,floatfix,longbibliography]{revtex4-2}\usepackage{graphicx} 
\usepackage{amsmath}
\usepackage{amssymb}
\usepackage{amsbsy}
\usepackage{graphicx}
\usepackage{color}
\usepackage{hyperref} 
\usepackage{enumerate}
\usepackage{mathtools}
\usepackage{lipsum}

\begin{document}
\title{Quasi-crystalline order in vibrated granular matter}
\author{A. Plati}
\affiliation{Universit\'e Paris-Saclay, CNRS, Laboratoire de Physique des Solides, 91405 Orsay, France}
\author{R. Maire}
\affiliation{Universit\'e Paris-Saclay, CNRS, Laboratoire de Physique des Solides, 91405 Orsay, France}
\author{E. Fayen}
\affiliation{Universit\'e Paris-Saclay, CNRS, Laboratoire de Physique des Solides, 91405 Orsay, France}
\author{ F. Boulogne}
\affiliation{Universit\'e Paris-Saclay, CNRS, Laboratoire de Physique des Solides, 91405 Orsay, France}
\author{ F. Restagno}
\affiliation{Universit\'e Paris-Saclay, CNRS, Laboratoire de Physique des Solides, 91405 Orsay, France}
\author{ F. Smallenburg}
\affiliation{Universit\'e Paris-Saclay, CNRS, Laboratoire de Physique des Solides, 91405 Orsay, France}
\author{G. Foffi}
\affiliation{Universit\'e Paris-Saclay, CNRS, Laboratoire de Physique des Solides, 91405 Orsay, France}
\begin{abstract}


Quasi-crystals are aperiodic structures that present crystallographic properties which are not compatible with that of a single unit cell. Their revolutionary discovery in a metallic alloy, more than  four decades ago, has required a full reconsideration of what we defined as a crystal structure. Surprisingly, quasi-crystalline structures have been discovered also at much larger length scales in different microscopic systems for which the size of elementary building blocks ranges between the nanometric to the micrometric scale. Here, we report the first experimental observation of spontaneous quasi-crystal self-assembly at the millimetric scale. This result is obtained  in a  fully athermal system of  macroscopic spherical grains vibrated on a substrate. Starting from a liquid-like disordered phase, the grains begin to locally arrange into three types of squared and triangular tiles that eventually align, forming  8-fold symmetric quasi-crystal that has been predicted in simulation but not yet observed experimentally in non-atomic systems.  These results are not only the proof of a novel route to spontaneously assemble quasi-crystals but are of fundamental interest for the connection between equilibrium and non-equilibrium statistical physics.



\end{abstract}
\maketitle

In 1982, Shechtman discovered the first alloy with a diffraction pattern for which the Bragg peaks showed a symmetry that is forbidden by crystallography in  periodic  solids \cite{Shechtman1985}. This discovery was initially met with resistance: the existence of structures in which atoms can  be arranged in spatial structures which lack long-range periodicity, while still preserving sufficient long-range order to generate discrete Bragg peaks clashed with the elegant picture of crystals as consisting of a repeating unit cell. Nonetheless, eventually  materials with this property, which were called quasi-crystals (QCs)~\cite{Levine84}, changed the way in which scientists interpret the crystal state, by disentangling the concept of order from the concept of periodicity, to the point where the very definition of crystals had to be changed to include aperiodic structures \cite{IUCR92}.

In the following years, quasi-crystalline structures have been observed in several artificial alloys (see Refs. ~\onlinecite{divincenzo1999quasicrystals, Ranganathan1991} for a review)   and were even discovered in a natural occurring mineral, Icosahedrite~\cite{Bindi_2009}, of probable extraterrestrial origin~\cite{Bindi2012}.  More recently, quasi-crystals have been also observed at much larger length scales in a wide range of soft matter systems \cite{Hayashida2007,barkan2011stability,Talapin2009,Takano2005,Zeng2004,Zhang2012,Lifshitz2007,Lee2010,Wasio2014,Förster2013} revealing promising optical properties for next-generation photonic devices \cite{Jin1999,Zoorob2000}. In soft quasi-crystals~\cite{Lifshitz2007}, two fundamental questions arise: i) understanding up to which length scales we can observe spontaneous quasi-crystalline order and ii) identifying the key dynamical and interaction properties required for a soft-matter system to form a quasi-crystal. 
For the former, the self-assembly of QCs has been observed on length scales that are related to the nature of the elementary building blocks that range from macromolecular structures~\cite{Liu2022} and nanoparticles~\cite{Talapin2009} to polymers aggregates made of micelles~\cite{Zeng2004,Fischer2011}, passing by polymers~\cite{Hayashida2007,Zhang2012,Lee2010,Wasio2014} and thin films~\cite{Förster2013}. To our knowledge, the largest soft-matter quasi-crystals found to self-assemble consisted of micrometer-size micelles~\cite{Fischer2011, Xiao2012}. 

The second question is of a more fundamental nature and has been explored extensively using numerical simulations. In particular, simple coarse-grained models of interacting particles make it possible to simulate systems large enough to display aperiodicity and to pin down its origin,
from anisotropic repulsive~\cite{HajiAkbari2009} and attractive~\cite{Noya2021} particles to simple isotropic potentials~\cite{widom1987, Dotera2014, malescio2022self} and hard spheres~\cite{Fayen2022}.
Interestingly, one of the simplest systems leading to quasi-crystal formation in silico is a simple two-dimensional binary mixture of non-additive hard disks undergoing equilibrium dynamics~\cite{Fayen2020,Fayen2022}. This proved that, in the right geometrical conditions, quasi-crystalline order can emerge from a purely entropy-driven self-assembly process. To our knowledge, quasi-crystal self-assembly in dissipative non-equilibrium systems where energy is constantly supplied from the environment and internally dissipated (such as active or granular matter) is still unexplored. 

A natural avenue to explore quasi-crystal formation beyond colloidal scale and thermodynamic equilibrium is to use granular matter, which has proven itself to be an ideal playground for the exploration of non-equilibrium phenomena over the last three decades. Depending on how a granular system is confined and the imposed external driving, it may exhibit either a fluid-like or a solid-like behaviour \cite{Jaeger96,AndreottiBook} and can undergo a variety of phase transitions \cite{Eshuis2007,Olafsen98,Reis2006,Aaranson2006,Panaitescu2012,Komatsu2015}. However, spontaneous quasi-crystal formation in systems driven out of thermodynamic equilibrium has not yet been observed in either experiments or simulations.

In this paper, we report the experimental and numerical observation of quasi-crystalline order in a binary mixture of millimeter-size spherical grains vibrated on a substrate.  
Our findings demonstrate that quasi-crystals can be formed also by macroscopic particles  much beyond the  scale at which thermal agitation plays a role. Our system is indeed intrinsically out of equilibrium  due to dissipation arising from frictional forces  and  energy injection due to external driving. 

\begin{figure*}
\centering
\includegraphics[width=0.93\textwidth,clip=true]{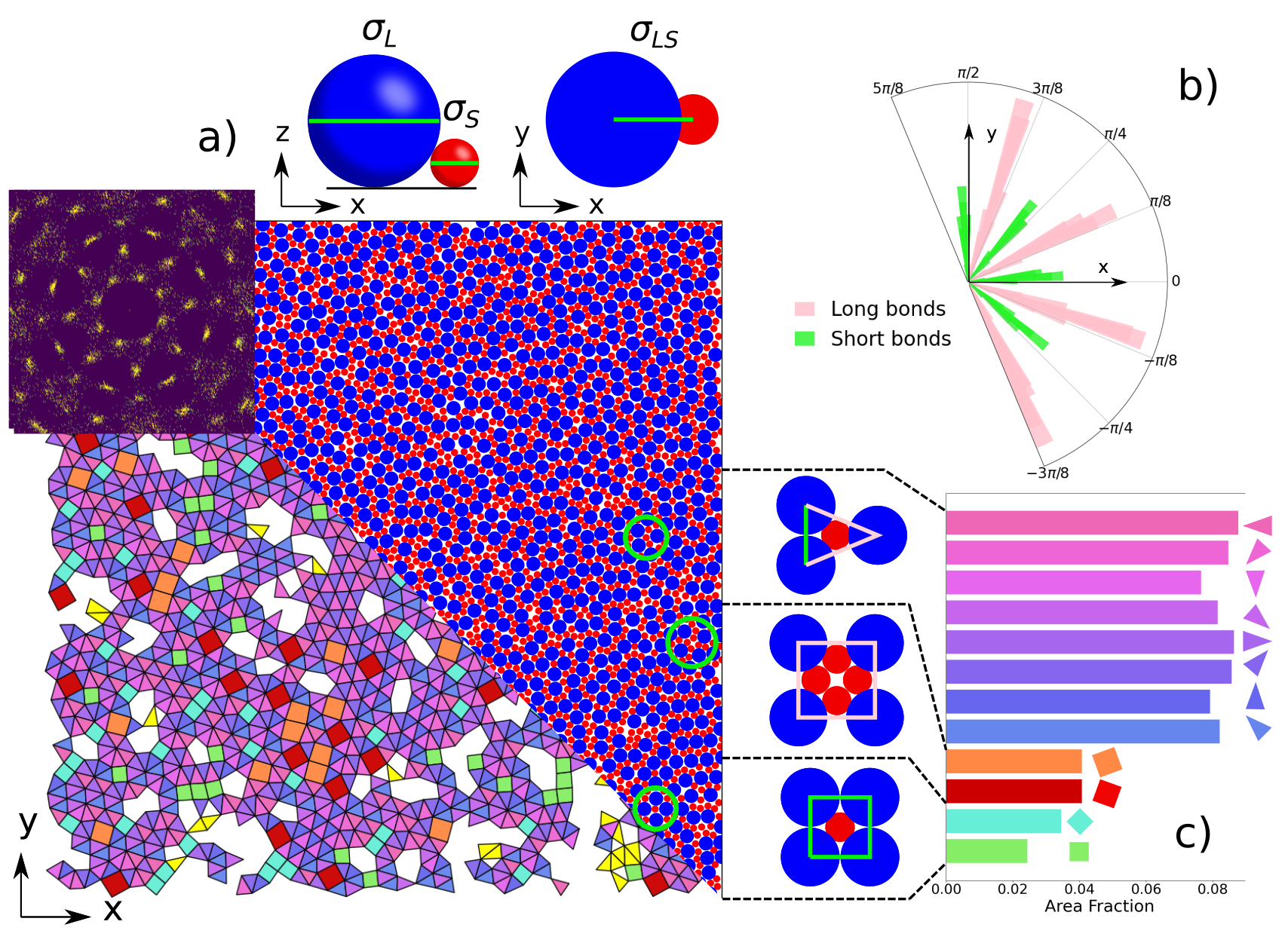}
\caption{
Numerical results obtained with EDMD of the collisional model defined in Eqs. \ref{eq: collRule}. Here $\{q,x_S,\phi\}=\{0.5,0.68,0.85\}$, $N_S+N_L=5000$, $\alpha=0.95$, $\Delta=0.02$. a) Sketch of the granular non-additive hard-core interactions, final configuration, reconstructed tiling and structure factor. b) Bond orientation histogram. c) Histogram of the area fraction occupied by tiles of different types and orientations. The color of the tiles is chosen according to their orientation but we also consider defected regions which are not covered by the correct square-triangle tiling (white areas) and tiles with ambiguous orientations or misaligned with the dominant set of bond directions (yellow areas). Results are obtained with a $5.3\times 10^{10}$ collision-long simulation.  \label{fig:Fig1}} 
\end{figure*}

In the following, we first discuss granular quasi-crystal formation in numerical simulations of a coarse-grained collisional model. 
We then report the main result of this paper namely the experimental self-assembly of a granular quasi-crystal. 

In both simulations and experiments, we consider a binary mixture of spherical grains ($N_S$ with diameter $\sigma_S$ and $N_L$ with diameter $\sigma_L$) lying on a horizontal substrate. In order to characterize the geometrical properties of such a mixture, we use the following dimensionless parameters: the size ratio $q=\sigma_S/\sigma_L$, the fraction of small grains $x_S=N_S/(N_S+N_L)$ and the area fraction $\phi=(N_S\sigma_S^2+N_L\sigma_L^2)\pi/4L^2$, where $L$ is the side of the box. As shown in Fig. \ref{fig:Fig1}a, spheres on a substrate can be mapped onto disks thanks to the introduction of a non-additivity parameter $\delta=2\sqrt{q}(1+q)-1$ and letting them interact at a distance smaller than $\sigma_{LS}=\frac{1}{2}(\sigma_S+\sigma_L)(1+\delta)$, note that $-1<\delta<0$. This is the main idea underlying the effective 2D model considered in a previous numerical study focusing on elastic non-additive hard disks following dynamics at thermodynamic equilibrium \cite{Fayen2022}.
This previous study revealed that, for sufficiently high area fractions and depending on the specific $\{q,x_S\}$ combination, one can observe the self-assembly of different crystals including a 12-fold and a 8-fold symmetric quasi-crystal. These results are in agreement with the fact that for conservative dynamics, geometrical constraints and hard-core interactions can be minimal ingredients for thermodynamic stability of quasi-crystalline structures \cite{haji2009disordered,Fayen2023}.
In this work, we focus our attention on the quasi-crystalline 8-fold symmetric phase (QC8), which was observed to self-assemble significantly more rapidly than the dodecagonal quasi-crystal \cite{Fayen2022}.

The collisional model we use in our simulations represents the athermal/non-equilibrium extension of the one considered in~\cite{Fayen2022}. It 
has been shown to embody dissipative and forcing mechanisms of experimental systems where spherical grains are placed on vertically vibrating horizontal substrates~\cite{Brito2013,Brito2020,Garzo2021}. In these systems, energy is injected along the vertical direction through grain-substrate collisions and then transferred to the horizontal ones through grain-grain collisions with an efficiency that depends on the impact kinematics.
In the model the dynamics is fully 2D: there is no vibrating plate since its effect is coarse-grained out thanks to the introduction of instantaneous grain-grain collisions which take into account both energy injection and dissipation. In this model, a binary collision between grains of mass $m_i$, $m_j$ obey the following rule for the velocity ($\mathbf v_i, \mathbf v_j$) update:
\begin{equation}
    \begin{split}
        \mathbf v_i'&= \mathbf v_i + \frac{m_j(1+\alpha)}{m_i+m_j}(\mathbf v_{ij}\cdot \boldsymbol{\hat{\sigma}}_{ij})\boldsymbol{\hat{\sigma}}_{ij} + \frac{2 m_j}{m_i+m_j}\Delta \boldsymbol{\hat{\sigma}}_{ij} \\
        \mathbf v_j'&= \mathbf v_j - \frac{m_i (1+\alpha)}{m_i+m_j}(\mathbf v_{ij}\cdot \boldsymbol{\hat{\sigma}}_{ij})\boldsymbol{\hat{\sigma}}_{ij} - \frac{2 m_i}{m_i+m_j}\Delta \boldsymbol{\hat{\sigma}}_{ij} ,
    \end{split}
    \label{eq: collRule}
\end{equation}
where the primed letters refer to post-collisional variables, and $\boldsymbol{\hat{\sigma}}_{ij}$ and $\mathbf v_{ij}$ are respectively the unit vector joining particles $i$ and $j$ and the relative velocity between them. The parameter  
$\alpha$ is  the coefficient of restitution  that embodies, for $0\leq\alpha< 1$, the dissipative nature of the collision. The last term in the Eqs. \ref{eq: collRule}, is responsible, via the parameter  $\Delta$, for the velocity gain arising from the non-planar collisions which are coarse-grained out in the effective 2D description (see SI for a more detailed explanation).
By computing the energy change in a collision~\cite{Brito2013}, it is possible to see that, depending on the impact kinematics, one can have conditions in which the total energy decreases or increases. In this simple granular model, the limit to the equilibrium conservative case is recovered by setting $\alpha=1$ and $\Delta=0$.
 
A granular system cannot attain thermodynamic equilibrium but it can reach a non-equilibrium steady state thanks to a balance between injected and dissipated energy. Nevertheless, one can often identify a conservative system with the same geometrical properties and consider it as an  equilibrium counterpart. In this perspective, non-additive hard disks in the conservative limit ($\alpha=1$ and $\Delta=0$) represent the equilibrium version of our granular system. 
Of course, given the dissipative/athermal nature of the dynamics, one cannot expect theoretical or numerical predictions for the equilibrium counterpart to hold in the granular case.    
However, in some specific conditions, vibrated granular materials have been shown to exhibit an equilibrium-like phenomenology, such as in the case of tracer diffusion in granular gases \cite{DAnna2003,Sarracino2010} or hexagonal crystal formation in monodisperse granular layers vibrated on a substrate \cite{Olafsen98,Reis2006,Komatsu2015}. The latter phenomenon is particularly relevant for our study since many aspects of the equivalent elastic hard-sphere crystallization have been observed also in the liquid-solid granular phase transition. The main challenging aspect for the observation of an equilibrium-like behaviour in our system is given by size polydispersity because it usually triggers non-equilibrium effects in vibrated granular materials. Size segregation \cite{Aaranson2006,Aumaitre2001,Baldassarri2015,Kudrolli2004} and violation of energy equipartition \cite{Klebert2002}, are two examples of that.  Thus, an important question underlying the approach we propose is the following: can we tune the non-equilibrium parameters i.e. related to forcing/dissipative mechanisms of the granular system, such that it self-assembles into the quasi-crystalline structures that have been observed in the equilibrium counterpart? 

Results obtained through event-driven molecular dynamics (EDMD) simulations of the model described by Eqs. \ref{eq: collRule}  are reported in Fig. \ref{fig:Fig1}a. There we show the last snapshot of a simulation and the associated scattering pattern computed from the large grain positions. The final self-assembled structure exhibits no periodic order but the scattering peaks reveal an underlying 8-fold symmetry. This particular symmetry is forbidden by ordinary crystallographic order based on repeated translations of a single unit cell. Indeed, in our system, the final structure can be decomposed into a combination of three different tiles each one appearing with different orientations. In Fig. \ref{fig:Fig1}a, we also highlight how small and large grains combine to form such tiles: we have i) small squares made of four large grains surrounding one small grain, ii) isosceles triangles made of three large grains surrounding one small grain and iii) large squares made of four large grains surrounding a square of small grains. The sides of the tiles coincide with bonds between large grain nearest neighbours and we can identify long and short bonds. The former outlines large square sides and triangle legs, the latter forms small square sides and triangle bases.  In addition to the scattering pattern, another piece of evidence of the 8-fold symmetry is given by the histogram of bond orientations over the entire system (Fig. \ref{fig:Fig1}b). Here we can clearly see that short and long bonds are not uniformly distributed but are aligned along eight specific directions. These dominant directions are spontaneously selected among a continuum since PBC do not impose preferred orientations. Within a specific 8-fold set of bond directions, small and large squared tiles can only appear with two specific orientations while triangular tiles can lay along eight ones. One can then classify all the tiles in the system (see Methods) as shown in Fig. \ref{fig:Fig1}c and reconstruct the overall tiling of the plane (bottom-left side of Fig. \ref{fig:Fig1}a).  The observed square-triangle tiling can cover aperiodically an infinite plane with long-range 8-fold orientational order. What we observe in our simulations is a finite portion of such an infinite quasi-crystal, the fact that tiles with the same form but with different orientations cover a similar area fraction is coherent with this picture \cite{Fayen2022}. Additional EDMD simulations varying the geometrical parameters $\{\phi, x_S, q\}$, the non-equilibrium ones $\{\alpha,\Delta\}$ and considering different system sizes confirmed the robustness of QC8 formation for this model (see SI).

We now turn our attention to the experimental realization. Our setup consists of non-magnetic steel spherical grains confined in a quasi-2D squared container (height $h$ and width $L\gg h$) which is vertically vibrated by an electrodynamic shaker following a signal $z_p(t)$. The evolution of the system is followed by a camera that allows to detect the horizontal position of the grains ($xy$ coordinates) as a function of time. The relevant set of explored geometrical parameters $\{q,x_S,\phi\}$ for the experiments is chosen close to those used in the simulation model. The realistic system  eventually attains a non-equilibrium steady state whose dynamical properties sensitively depend on the driving force. To tune this, we performed preliminary numerical simulations of the setup implemented through the Discrete Element Method (DEM) \cite{Cundall79,PoeschelBook}. Such simulations implement granular dynamics by means of accurate contact models
which allow for studying in-silico prototypes of realistic setups (see SI). From this analysis and subsequent experimental tests (see SI), we found that to observe QC8 self-assembly one generally needs sufficiently strong vibrations to allow for an efficient vertical-to-horizontal energy transfer but not so strong that grains pile up on each other (which alters the effective 2D packing fraction of the system).


\begin{figure}
\centering
\includegraphics[width=0.99\columnwidth,clip=true]{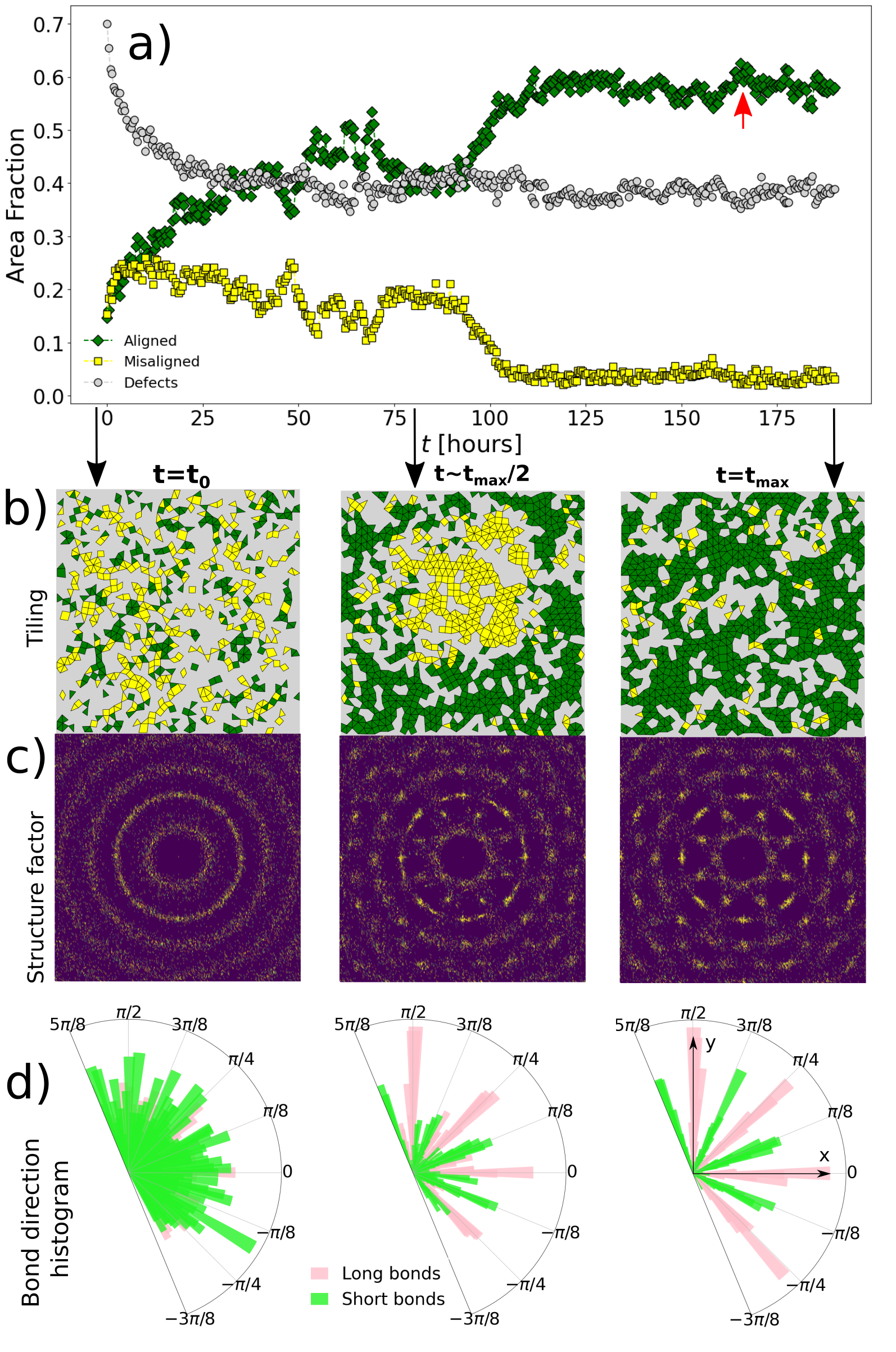}
\caption{ Experimental results for $\{q,x_S,\phi\}=\{0.5,0.675,0.848\}$, $N_S+N_L=3840$, $\sigma_L=2.381$ mm, $z_p(t)=A\sin(2\pi f t)$ where $A=22$ $\mu m$ and $f=120$ Hz. Data from real-time measurements during QC8 formation (see also supplemental video). a) Area fraction of tiles correctly aligned according to the dominant orientational order, misaligned tiles and defects. Reconstructed tiling (b), structure factor (c) and bond orientation histogram for early, intermediate and late times. Yellow tiles are misaligned with the final long-range order, green ones are in agreement with it. We note that the latter have long bonds laying on the $xy$ axis. See also Movie 1. \label{fig:Fig2}}
\end{figure}


\begin{figure*}
\centering
\includegraphics[width=0.99\textwidth,clip=true]{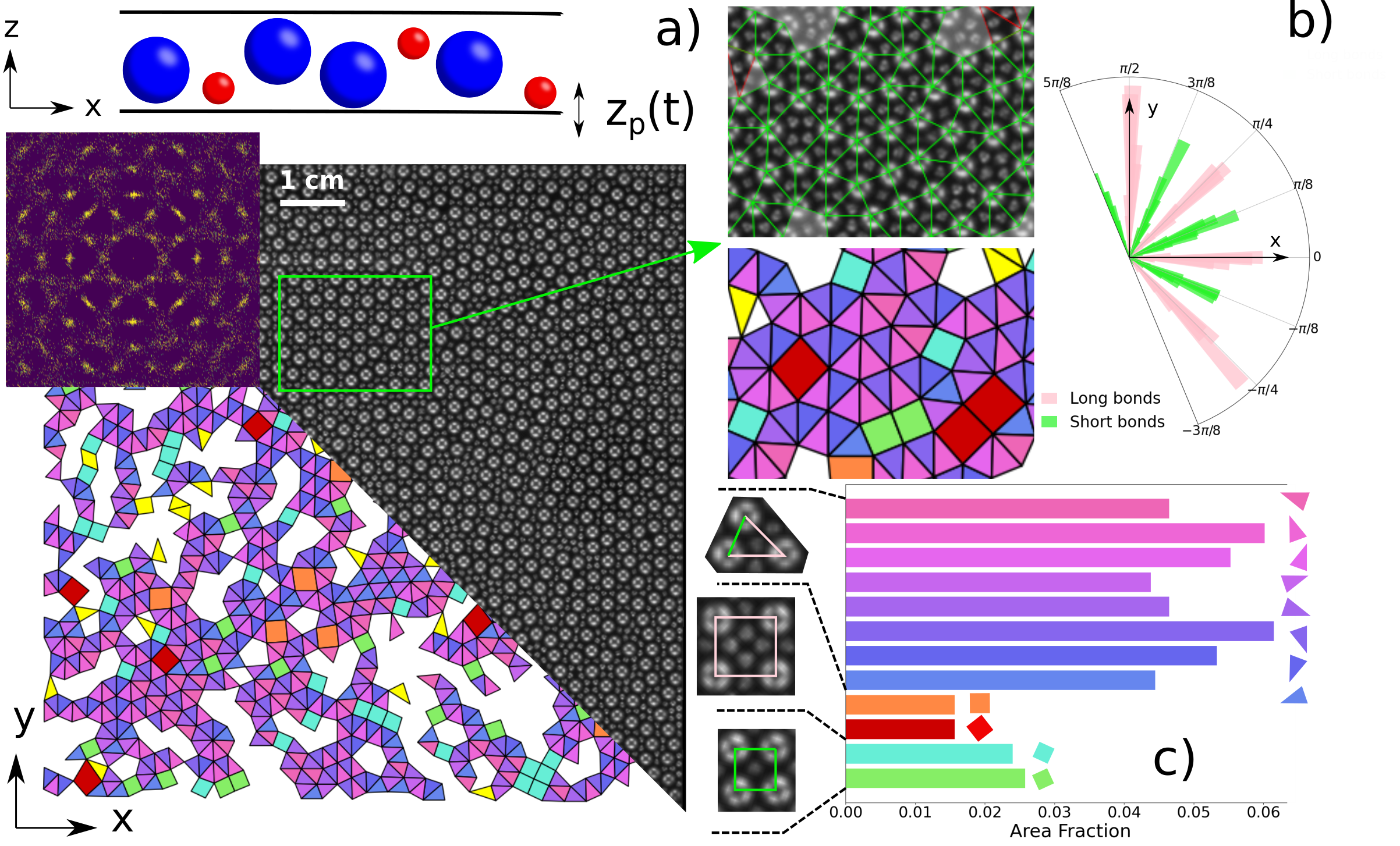}
\caption{
 a) Configuration, structure factor, reconstructed tiling obtained after $\sim 170$ hours of the experimental run already shown in Fig. \ref{fig:Fig2}. The specific snapshot shown is the one with the highest QC8 area fraction (red arrow in Fig. \ref{fig:Fig2}). We also show a sketch of the $xz$ projection of the setup and a comparison between real space tiling and the reconstructed one in a zoomed region on the sample. b) Bond orientation histogram. c) Tile area fraction histogram with snapshot of single granular tiles forming the quasi-crystal. \label{fig:Fig3}}  
\end{figure*}

For suitable choices of the driving parameters, our experimental system indeed spontaneously forms the 8-fold quasi-crystal. The dynamics of this process is shown in Fig. \ref{fig:Fig2}a (see also supplemental video) where we plot the occupied area fraction as a function of time for three different groups of tiles: the ones oriented according to the specific 8-fold set of directions which dominates at the end of the experiment (green), misaligned tiles (yellow) and defects (grey).
These results are obtained from a seven-day-long experiment with sinusoidal vibrations and quantify the two distinct processes contributing to the QC8 self-assembly: single tile formation and tile orientational ordering. We note that both aligned and misaligned tiles increase monotonously at very short times. After that, their evolution becomes non-monotonous: aligned tiles exhibit an increase interrupted by sudden drops while misaligned tiles exhibit a decrease interrupted by sudden growths. Both of them eventually reach a final plateau. Such behaviour reveals that, once tiles are formed and locally ordered, they still need substantial rearrangements in order to form larger quasi-crystalline domains and this makes the QC formation extremely slow. 

From the evolution of reconstructed tilings shown in Fig. \ref{fig:Fig2}b, we observe
that QC8 with the dominant 8-fold symmetry is growing from the boundaries toward the bulk. During this process, it is possible to observe the coexistence between different misaligned QC domains ($t\sim t_{\text{max}}/2$).
In Fig. \ref{fig:Fig2}c, we also show the evolution of the structure factor: the initial configuration shows the typical rings of a liquid-like structure, the middle one exhibits blurred peaks originating from the coexistence of multiply oriented QC domains while the last one presents sharper peaks highlighting long-range order according to a dominant set of 8 directions.  Finally, from the evolution of the bond orientation histogram (Fig. \ref{fig:Fig2}c), we can see that the
most favoured set of orientations is the one with large tile sides aligned with the hard walls. Repetitions of the experiment revealed that this is a reproducible feature.
Indeed, an important difference with respect to the numerical case discussed above is represented by the hard horizontal walls which break the orientational symmetry. The enforced grain alignment at the $xy$ boundaries favours two specific sets of 8-fold symmetric directions: one with short bonds and one with long bonds aligned with walls. The dominance of the latter can be explained considering that the forming quasi-crystalline structure requires much more long bonds than short ones.

An interesting picture  emerges:  tiles of the desired shapes form very rapidly, probably as a consequence of their efficient local packing. However, global alignment requires much more collective rearrangements that appear to be rare events.
It is important to point out that real-time measurements of forming quasi-crystalline structures are extremely rare in experiments. Generally, the techniques used for QC self-assembly (e.g. evaporation of nanoparticle solutions) are not compatible with the observation of the dynamics during the ordering process but only allows to analyze the final structure. The experimental study of system configurations over time is then another important novelty of macroscopic quasi-crystals since it can shed light on unexplored dynamical properties of QC self-assembly.

In Fig. \ref{fig:Fig3}a, we show the spatial configuration, the related scattering pattern, and the reconstructed tiling obtained from the experimental configuration with the largest quasi-crystalline area fraction (see the red arrow in Fig. \ref{fig:Fig2}a). On the right side of the same figure, we report the histograms of bond orientation (b) and tile area fraction (c). We note that the key properties of the quasi-crystalline structure observed in EDMD simulations are found also in the experiment: we have the same 8-fold long-range orientational order, confirmed by the scattering intensity, and the same square-triangle tiling. By comparing Fig. \ref{fig:Fig1} and \ref{fig:Fig3}, we realize that the main difference between EDMD simulations and experiments is the presence of larger defects in the latter. It is difficult to pin down the origin of this difference as the $z$-to-$xy$ energy transfer mechanism of the real experimental model  depends, in a highly non-trivial way, on the dissipative properties of the materials and the plate roughness, while the energy injection in the EDMD simulations is fully described by a simple collision rule that depends only on two parameters. We argue that larger defects are also the reason why we observe lower large-square area fraction in the experiment with respect to EDMD (compare Fig. \ref{fig:Fig1}c and \ref{fig:Fig3}c). In fact, large squares are favoured by a large number of small grains \cite{Fayen2022}, but many of them are stuck in defects that appear mainly as clusters of small grains. Despite these differences, we find it remarkable that the EDMD simulations effectively capture the essence of non-equilibrium self-assembly.  Moreover, this computational scheme is considerably simpler in comparison to DEM simulations and could serve as a vital tool in further explorations of this phenomenon.

In order to test robustness and reproducibility of the discussed results we repeated the experiment for different initial random configurations, different drivings (sawtooth vibrations) and slightly shifted state points. As reported in the SI, the emergence of quasi-crystalline order proved to be a quite robust feature of this granular system.

More than 40 years ago, the quasi-crystal self-assembly paradigm emerged in  purely atomistic systems but was later extended to the nanometric and micrometric scales typical of soft matter. Here, we have shown that this idea can be extended to much larger length scales in vibrofluidized granular systems. Our study reports the first observation of quasi-crystalline order in a physical system undergoing athermal dynamics in the visible scale where real-time measurements of system configuration can be performed during the self-assembly process. The passage from the micrometer to the millimeter scale is not a trivial one, since in the latter the grains do not undergo thermal agitation. In the seminal work by Edward~\cite{EDWARDS19891080}, entropic reasoning was employed to elucidate the behavior of jammed granular systems (or "powders" as termed by Edwards). However, such an approach hinges on the presumption of a jammed state, an assumption not applicable to our vibrofluidized systems. An apt theoretical framework should ideally extend the equilibrium vibrational entropy, which is instrumental in stabilizing QC~\cite{Fayen2023}, to non-equilibrium conditions, perhaps drawing upon kinetic theories. Despite this fundamental difference, quasi-crystalline order emerges in the same conditions  as predicted at equilibrium, something that could not be trivially expected.

\section*{Acknoledgments}
We thank Andrea Puglisi and Andrea Gnoli for their invaluable help in setting up this project and for their comments on the manuscript. We also thank Marianne Impéror-Clerc, Laura Filion,  Anuradha Jagannathan and Franscesco Sciortino for carefully reading and commenting on our paper and Stéphane Cabaret for the design of the quasi-2D cell.
This work has been done with the support of Investissements d'Avenir of LabEx PALM (Grant No. ANR-10-LABX-0039-PALM) and of the Agence Nationale de la
Recherche (ANR), grant ANR-18-CE09-0025. 

\section*{Data availability}

Data sets  generated during this study  as well as all custom computer codes and scripts used to generate them  are available from the corresponding author on reasonable request.

\bibliography{biblioQC}

\end{document}